\documentclass[aps,prd,preprint,groupedaddress,nofootinbib,floatfix]{revtex4-1}
\pdfoutput=1
\usepackage{graphicx}
\usepackage{epsfig}
\usepackage{amsmath,amssymb,hyperref,enumerate}
\usepackage{slashed}
\usepackage{amsfonts}
\usepackage{url}
\usepackage{color}
\usepackage{bm}
\usepackage{subfig}
\hypersetup{colorlinks,citecolor= black,linkcolor= black}
\definecolor{nicered}{rgb}{0.7,0.1,0.1}
\definecolor{nicegreen}{rgb}{0.1,0.5,0.1}
\usepackage{setspace}

\allowdisplaybreaks

\def\({\left(}
\def\){\right)}
\def\[{\left[}
\def\]{\right]}

\usepackage[margin=1.in]{geometry}

 \usepackage{color}    
 \usepackage{graphicx} 
\usepackage{amsmath}

\begin{document}

 \title {Bounds on Neutron- Mirror Neutron Mixing from Pulsar Timings and Gravitational Waves Detections}
\author{  I. Goldman$^{b,a}$}\author{ R. N. Mohapatra$^c$}, \author{S. Nussinov$^{a}$,  }
\affiliation{ 
$^a$School of Physics and Astronomy, Tel Aviv University, Tel Aviv, Israel  }

\affiliation{$^b$Department of Physics,  Afeka Tel Aviv   Engineering College, Tel Aviv, Israel  }
  
\affiliation{ $^c$Maryland Center for Fundamental Physics and Department of Physics, University of Maryland, College Park, USA  }

%\affiliation{ $^d$Schmid College of Science,Chapman University, Orange, CA 92866, USA  }

\begin{abstract}

The mass loss in putative neutron star  to mixed neutron - mirror
neutron star transition implies a significant change of orbital
period. The precise constancy of the latter can restrict scenarios
recently suggested where neutron to mirror neutron mixing occurring in
neutron stars, transforms them into mixed stars helping explain the
narrow mass distribution observed for pulsars in binary systems. The observation of a very old millisecond pulsar with a mass of 2 solar masses is an additional strong constraint on the above transition.We also note that the observed gravitational waves signals from neutron-neutron stars merger constrain the neutron to mirror neutron transitions inside neutron stars. 
These considerations exclude a large region in the $\epsilon'$,
$\delta m'$  plane of the neutron-mirror neutron mixing and mass
difference.

\end{abstract}

 \maketitle

\section{  Introduction}

The possibility that there may be a mirror sector of the standard model  with identical particle content and gauge symmetry as the standard model ~\cite{mirror1} has received a great deal of attention. In particular, the dark matter of the universe as well as any sterile neutrinos may be the lightest baryon (or atom) and the light neutrinos of the mirror sector respectively. Prior to symmetry breaking, these models assume the existence of $Z_2$ (mirror) symmetry between the two sectors which keeps the same number of parameters even though the number of particles is doubled making this a very economical scenario of beyond the standard model physics.  Two realizations of these theories have been extensively discussed: one where mirror symmetry is respected by the electroweak vacuum expectation values (vevs) ~\cite{mirror3} and another where the mirror symmetry is broken via different electroweak scales in the two sectors~\cite{mirror2}. In this paper, we will focus on the first class of models where $v_{wk}=v'_{wk}$ (where $v_{wk}$ and $v'_{wk}$ denote the vev of the standard model Higgs field and the mirror SM Higgs field; we will use prime to denote the entities in the mirror sector) and we have identical microphysics and the same fermion masses on both sides. There is, however, a generic problem with both these scenarios in that the extra neutrinos ($\nu'$) and photon ($\gamma'$) from the mirror sector contribute  too much to the number of degrees of freedom at the BBN epoch, destroying the success of the big bang nucleosynthesis predictions. A cure for this is  to assume a breaking of the $Z_2$ symmetry in the early universe so as to have asymmetric inflationary reheating in the two sectors resulting in a lower reheat temperature ($T'$)  of the mirror sector compared to that ($T$) of the visible one~\cite{asym}. This breaking eventually trickles down to the low energies leading in general to a splitting the mirror and visible fermion masses~\cite{MN}. The above mentioned symmetric picture could however remain almost
exact if the asymmetric inflation picture is carefully chosen~\cite{MN}. Cosmology of such scenarios have been discussed in ~\cite{bere,foot} .%The dark matter  in both scenarios can be the mirror neutron (denoted by $n'$) with same mass as the known neutron ($n$) a possibility that has been extensively studied in recent literature 
 
 An interesting new phenomenon is possible in almost exact mirror models, if there are interactions mixing the neutron with the mirror neutron state ( denoted by $\epsilon_{n-n'}\equiv \epsilon'$), then one can expect $n\to n'$ oscillation to take place in the laboratory~\cite{BB} and indeed there are ongoing and already completed searches for such oscillations~\cite{expt,yuri} at various neutron facilities. We note that  $n-n'$ oscillation is very similar to neutron-anti-neutron oscillation suggested very early~\cite{nnbar} and also discussed extensively in literature~\cite{review}. For $n-\bar{n}$ oscillation to occur, one needs a $\Delta B=2$ interaction that leads to a mixing mass $\epsilon_{n-\bar{n}}$ between the $n$ and $\bar{n}$ similar to $\epsilon_{n-n'}$. %The $\epsilon_{n\bar{n}}$ acts like a Majorana component to the neutron mass matrix. 
There is however a major difference between these  two oscillations. Whereas the desired mass equality of neutron and anti-neutron,is guaranteed  by the
CPT theorem,  known to be an exact consequence of local, Lorentz invariant Quantum Field Theories, the mass degeneracy between $n$ and $n'$ oscillation requires an almost exact $Z_2$ symmetry, which as discussed above, needs to be a weakly broken symmetry to satisfy BBN constraints. %However, as noted, for cosmological consistency of the theory this symmetry needs to be broken.
%underlies  $ n \to \bar{n}$  oscillations (9) (10) . These
%oscillations were suggested before the  $(n\to n')$  oscillations ( 10, 11)
%which require a very high degree of mirror symmetry to  ensure a tiny
%$m(n)-m(n')$ difference.
Stringent upper bounds on $\epsilon_{ n\bar{n}}$ , the neutron
anti-neutron mixings  have been obtained by searching for $ n \bar{n}$
conversion in magnetically shielded neutron beams~\cite{baldo}  and in nuclei~\cite{superk}. The
subsequent annihilation of the generated $\bar{n}$  can be readily
identified in both cases. The $(n\to \bar{n})$   transitions in nuclei are highly suppressed by the
ratio: $\epsilon^2_{n\bar{n}}/{(E-\bar{E})^2}$ with
$\epsilon_{n\bar{n}}= 1/ \tau_{n\bar{n}}$ the off diagonal $\Delta B=2$
element of the $2 \times 2$ energy -mass matrix for the  $ n  \bar{n}$  
 system and
$E-\bar{E}$ the large  difference between the diagonal elements - the
energies  of the neutron and anti-neutron in the nucleus. Still the
large number of neutrons in the underground detectors compensates this enormous suppression and quite
remarkably the same upper limit:
\begin{eqnarray}
\epsilon_{n \bar{n}}= 1/{\tau_{n\bar{n}}} < 10^{-8}~ Sec^{-1} ~{\rm or}~ 10^{-23}~{\rm eV}
\end{eqnarray}
is obtained both by direct oscillation searches as well as by nuclear decay searches.

The upper bounds on $n\to n'$ , the analog transitions are far weaker. 
The $n'$ generated in neutron beams or bottles simply leave the system manifesting only in a
deficiency of neutrons beyond what is expected from the neutron decay
only~\cite{expt}.
\begin{eqnarray}
\epsilon_{n n'} =
\epsilon' \leq 10^{-16}~{\rm  eV}.
\end{eqnarray}
 The analysis is further complicated by the possible presence of
mirror magnetic fields which can suppress the transition rate and,
unlike ordinary B fields,  cannot be shielded. That allows for $\epsilon'$ values which are significantly higher.
Also unlike for
$n\to\bar{n}$, $n\to n'$ transitions inside nuclei are energetically
forbidden as the neutrons are bound (typically by $\sim 8$ MeV) and the
equal mass mirror neutrons with no (ordinary) nuclear interactions are
unbound.

Here we focus on the recent suggestion~\cite{mana} that there may be another manifestation of neutron to mirror neutron $(n\to n')$
transition with important implications i.e. it can occur in neutron stars~\cite{n*} and can convert  neutron stars to mixed
neutron + mirror neutron stars. This may help explain some peculiarities of
the mass distribution of pulsars.
The key to the new suggestion of \cite{mana} which we  briefly elaborate 
in the next section, is that such transitions are   allowed
in neutron stars even though they are forbidden in nuclei. Furthermore, an analog of the remarkable coincidence
between the nuclear and beam method bounds obtained in the case of $n\to\bar n$
 transitions repeats here with the neutron stars playing the
role of ``giant nuclei". Specifically sufficiently large, yet still
allowed values of $\epsilon'$ which are probed in beam experiments can convert an initial neutron star
to a maximally mixed-lower mass star on relevant time scales of $T \sim
10^6-10^{10}$ years.

It has been argued ~\cite{mana} that this will modify the mass distribution of the pulsars which has been
measured with precision for pulsars in binaries. The
effective softening of the combined equations of states leading to
smaller size and increased  gravitational binding, shifts the pulsar
masses towards lower values over the above  time scale $T$ making for a
better agreement with the observed mass distribution.

 Our main message,  presented in sec.III below,  is that this ingenious
and intriguing suggestion is strongly constrained in a model independent
way by precision measurements of the orbital periods of pulsars which
are members of binary systems. We also discuss the possible implications for the above scenario 
of the recent observation of gravity  waves emission in a binary neutron star merger.

  It is  well known~\cite{GN} that mass loss implies an increase of the
orbital period of the binary systems.% one of which is a neutron star. 
Our claim is that if the rates of conversion of
neutron stars to mixed stars are those required in the above 
scenario  (or even a thousand times slower)  the resulting period increase
exceed the maximum allowed by the observations. %This tends to exclude
%the intriguing scenario of Ref.~\cite{mana}  unless the conversion is very fast
%and is completed before the observations of the binary system are
%made.

In section IV,  we turn the argument around and use the putative neutron
star$\to$ mixed neutron--mirror neutron star transitions  to exclude a large
 range of  $\epsilon'$. This is
particularly relevant when the $n-n'$ mass differences $\delta m'$  are
significantly higher than $\epsilon'$ and the energy difference due to
magnetic fields.% \footnote{A case in hand is a recent paper~\cite{bere3}}. 
The ( almost) free $n\to n'$ oscillations in neutron
beams or bottles are than strongly suppressed and the astrophysical limits considered here
 are the {\it only} way to restrict $\epsilon'$.

In Sec V, we present some further speculative comments  and in sec. VI, we conclude with a summary of our results. In Appendix A,
we give our estimate of the $n\to n'$ transition rate inside a neutron star.

\section{Transition of neutron star to mixed  neutron --mirror neutron stars
induced by n-n$^\prime$ mixings}

 The suggestion that certain stars and neutron like stars in
particular , may be "mixed" --consisting of ordinary baryons and dark
matter particles- has been made some time ago~\cite{GMNRT, OTHER}.  The  present
authors together with Doris Rosenbaum and the late Vic Tepliz
considered \cite{GMNRT} such stars and solving the coupled TOV equations
derived allowed masses of stable mixed stars.  While we found, like
the authors of \cite{mana}, that admixing dark matter particles of equal or
heavier mass than neutrons pushes down the maximal allowed stellar
mass, having lighter -  say $m_{n'} = \frac{m_n}{2}$ - DM particles allows
exceeding the maximal neutron star mass of $\sim  2.3 M_{\odot} $ suggested by most
equations of states ( EoS) of ordinary nuclear/ quark matter\cite{Bedaque and  Steiner, Alford}.

%( THERE WAS YET ANOTHER ASTROPHYSICAL OBSERVATION MOTIVATING AN INNER,
%SMALLER PULSAR WITHIN A MORE MASSIVE BIGGER STAR -WHICH YOU ITZHAK
%ELABORATED . WHAT WAS IT ? AND SHOULD WE MENTION IT HERE? It was our unpublished manuscript (mia culpa) on the Ultra Luminous X ray source with luminosity corresponding to the Eddington value of about
%100 solar masses that was a pulsar and thus not a BH. Not sure if we gain much by mentioning it here. Somebody may use the idea and thus we should publish that paper before.)

 Two conceptual issues were encountered in previous mixed star discussions.

(i) There was no obvious mechanism for bringing together at some stage
the required roughly equal amounts of ordinary and dark matter into
the star. 

(ii)  The  EoS of the (self interacting) DM are unknown. In the broken
mirror models underlying  our above mentioned work~\cite{GMNRT},  we used scaling
with $m(n')$  to guess the latter from the EoS of ordinary matter.

  The new mixed star scenario suggested in ~\cite{mana} circumvents both
issues.

(i) The $n\to n'$ oscillations can generate (starting with a pure
ordinary neutron star)   a mixed $n-n'$ star and 
(ii) the exact mirror model used therein implies identical EoS of
mirror and ordinary matter.

The key to the arguments of ref.\cite{mana} is that $n\to n'$  transitions,
kinematically forbidden in nuclei, do occur in neutron stars,  causing
conversion of neutron stars to mixed ( $n-n'$) stars.  Furthermore these
transitions take relatively short   astrophysically relevant times of
   $ \sim 10\div 100$ Myr for values of the microscopic mixing
$ \epsilon'$ which are allowed by all other
terrestrial and cosmological limits . For completeness we will briefly
recap bellow some of the main aspects of this argument.

The reason why $n\to n'$ transitions do occur in neutron stars is that
the neutrons therein are mainly bound by gravity and {\it not} by
nuclear forces which at the large densities in the central regions of
the star may become repulsive. Thus, suppose that an  $n\to n'$
conversion occurred at some point in the star. Under the pressure which is proportional  roughly to the
energy density, a neighboring
neutron then rushes into the "hole"  generated
by the converted neutron,  gaining in the process kinetic energy %$p v \sim p (fm)^3$
 which is of  order of the Fermi energy $E_F$. Further energy is
gained when the produced $n'$ gravitates to the center of the star and a
surface neutron replaces the neutron which went into the above "hole".
 In reality we have continuous inward drifting of both mirror neutrons
$(n')$ and neutrons $(n)$ due to the ongoing $ n\to
n'$ transitions over  the whole volume of the star . This also causes the
star to shrink which further significantly increases the gravitational
binding.   Altogether this  decreases the stellar mass by about 20\%. %Altogether in each $ n\to
%n'$ transition we gain about twice
%the large Fermi energy E(F) of the disappeared neutron. (The initially
%low concentration implies a lower Fermi energies $E'(F)$ of the $n'$s for
%most of the duration of the stellar transition).
  Thus the transitions do occur, albeit with suppressed rates, similar
to $n\to \bar{n}$ in nuclei.
The estimate of the neutron to mirror neutron transition rate done in ref.\cite{mana} proceeds in two steps:

A) It is assumed that in each $n-n$ collision the fraction of $n'$ 
\begin{equation}
\label{P}
  P_{nn'}=\epsilon'^2/{[E_F-E'_F]^2}        
\end{equation}
 in the neutrons wave-function will materialize as mirror neutrons . and next

B)  The $nn$ collisions rate   $\Gamma_{nn}$ is independently given by:
\begin{eqnarray}
\label{col}
 \Gamma_{nn}= \sigma_{nn}v N
 \end{eqnarray}
where $\sigma_{nn}, v , N$     are the $nn$ collision cross-section, the
neutrons velocity and number density of neutrons in neutron star respectively . This rate was estimated  to be
\begin{eqnarray}\label{value}
\Gamma_{nn}(neutron~star)=\Gamma_{nn}(nuc) [N/{N_{nuc}}]^{4/3} \sim 10^{24} [N/{N_{nuc}}]^{4/3} ~sec^{-1}\sim 3.5\times 10^{24}~ sec^{-1}.
\end{eqnarray}
%with $N_{nuc}$ the nuclear number density and $N$ is the corresponding density in neutron star. We can reproduce the $\Gamma_{nn}(nuclear)\sim 10^{24}$ sec$^{-1}$ and  $N_{nuc}= 1.6\times 10^{38}$ cm$^{-3}$ if $ v\sim 0.3 c$  $\sigma_{nn}\sim 0.6\times 10^{-24}$ and $v\sim 0.3c$ The $4/3$ power above
%reflects the obvious $N$ factor in Eq.\ref{col} and another $N^{1/3}$ which is proportional to $p_F\propto v$. If $ E_F-E'_F\sim
%E_F\sim 40 MeV $ is  the (Fermi) energy difference in the denominator of
%Eq.\ref{P} and $N/N_{nuc}\sim 2$ in Eq.\ref{col}, we get 
finally leading to
\begin{equation} 
\label{rate}
 \Gamma(n \to n') =  \Gamma_{nn}  P_{nn'} \approx  10^{-6}  \left(\frac{\epsilon}{ 10^{-11}eV}\right)^2
{yr}^{-1}   
\end{equation}
% that was used in ref.\cite{mana}.
 The actual calculations of the evolution towards the mixed star
undertaken in Ref.\cite{mana} are rather complicated and far more elaborate.
One has to use the spatially varying density $n(r)$ and $n'(r)$ of the
neutrons and the mirror neutrons in the coupled TOV equations
describing the instantaneous hydrostatic equilibrium states of the
system. %{\color{red}(Shmuel- the density of a pure and also our mixed NS are  roughly   constant. Given that the conversion process time scale is enormously larger than the dynamical time scale of the neutron star there is hydrostatic equilibrium and thus the densities will be those of the static configuration)}.  
These equation and corresponding equilibrium states keep
changing in time as more and more neutrons are being converted to
mirror neutrons. %In turn the Fermi energies $E(F)$ and $E'(F)$ in the
%denominator of equation (\ref{P}) above keep getting closer as the number
%density $N'$ increases accelerating  the conversion rates in the inner
%regions of the star.  
These mirror neutrons keep drifting to the central region of the star generating
a region of a lowest energy completely mixed star. The equally mixed spherical region
gradually expands and eventually overtakes the whole star. 
Furthermore the calculations have to be done for
various  equations of states relating the energy density and pressure
of the ordinary nuclear matter (and of the mirror nuclear matter) in
the star. The scope of the present paper is much more limited and we will not delve into
these issues rather we use the final results obtained in Ref.\cite{mana}:

a) Over times of order 1 -- 100 Million years the neutron to mixed
star transition can be  largely completed for $\epsilon' = 10^{-11}  - 10^{-12}~eV$ and 

 b)  The original neutron star shrinks and if it was too massive to
start with, will collapse to a black hole. Otherwise it will wind up
at present  when being observed as a partially or almost completely
mixed star with a mass loss of 0.25-0.35 solar masses.

  Such collapses and mass decrease in the above described
transitions to a mixed star will then appreciably shift down and
{\it narrow} the mass distribution of pulsars in binaries making for a
better agreement with the observed distribution~\cite{mana}.%%%

\section{Astrophysical  Constraints on the Neutron Star to Mixed Star Transitions}

\subsection{limits From orbital period measurement in binary pulsars}
 We now describe our main result - limiting in a model independent way
the above scenario by the precise measurement of the  
orbital period time derivative of pulsars in binary systems.  Such period decrease
,expected whenever the mass of either member of the binaries
decreases, was first noted in 1925 by Jeans~\cite{jeans}. Using the constancy of
the product $M a $ with $M = M_1 +M_2$,  the total mass  and $a$, the
semi-major axis appearing in Kepler's law for the orbital period
$P_b= 2\pi\left(a^3/ {G_N M}\right)^{1/2}$, he obtained the simple expression for
the change of the period of the orbital binary motion $P_b$ as a
function of the mass change:
     \begin{eqnarray}  
     \frac{dP_b/dt}{P_b} = - 2  \frac{dM/dt}{M} .
     \end{eqnarray}
 $P_b$ is the period of binary orbital motion, as distinct from $P$ which denotes the period for individual pulsar.\footnote{The rate of change of the latter presumably due to magnetic breaking,  defines the estimated age of the pulsar via $P/\dot{P}$. The conversion of the original neutron star to a lower mass mixed
star  also decreases the radius of the stars leading to faster pulsar
rotations and shorter periods. Indeed In the framework of the ambitious nano gravity project~\cite{nano}
a remarkably accurate pulsar timing has been achieved . One example is
the " millisecond" pulsar PSR J1024-0719 for which a period change
of
 $dP/dt =( 1.8551 \pm 0.0001)\times 10^{-20}$
 was measured.
Unfortunately since here unlike for the binary systems we cannot
predict the rate of change due to " conventional" sources, the {\it full}
$(dP/dt)/P$   and { \it not} just the fractional error in it   can be
attributed to new physics and the ensuing bound does not improve.
}
 
 While Jeans envisioned mass loss due to the electromagnetic  radiation emitted by
the stars, the above relation holds for  {\it any}  form of  ``radiation"  e.g via neutrinos, axions, mirror photons
and mass loss incurred - so long as the emission from either member of
the binary is symmetric in its rest frame. {As noted however in ref.~\cite{GN}, if some fraction of the energy in the case of interest is
emitted  electromagnetically  - most likely as X rays in the case considered below- the extra
signatures would be very remarkable and more stringent limits
would obtain}.

As $\frac{dM/dt}{M}$ is negative for mass loss, the period increases and the average angular velocity
$\Omega_{orbital}= 2\pi/P_b$ decreases. This is not the case for gravitational radiation which is emitted from the rotating system as a whole, 
decreasing the orbit and {\it increasing} the $\Omega_{orbital}$.

This was used in the
famous inference of gravitational wave emission measurements from the period slowing down of PSR 1913+16 -the binary  Hulse
Taylor pulsar (age $1.1\times 10^8$ yrs). Two of us (I.G and S.N.) in ref~\cite{GN} suggested that the
same precise measurements can be used to constrain also the energy
loss due to continuous neutrino emission which might occur in  
models where neutron stars undergo internal changes over long times.
Using  the measured period slow-down, a total mass $M$ of about $3M_{\odot}$ 
and accounting for the gravitational wave mass loss which can be very accurately predicted, it was found that~\cite{GN}:
\begin{equation}
\label{limit}
\Delta M ~{yr}^{-1} < 3\times 10^{-12} M_{\odot} {yr}^{-1}  .       
\end{equation} 
 A somewhat stronger  limit  $\Delta M~{yr}^{-1} < (0.96-1.2)\times 10^{-12} M_{\odot} {yr}^{-1}$ can be derived from observations on PSRJ1952+2630.
  This is in a binary orbit with a $0.93-1.4 M_\odot$  white dwarf~\cite{Lazarus et al. 2014}  . 
 Its spin down age is $7.7 \times 10^7$ yr, the orbital period is $0.39$ days and during 800 days of follow-up the error on the period is $ 7\times 10^{-13}$ days. Thus
 \newpage
\begin{equation}
\label{J1952+2630}
|\frac{\dot P_b}{P_b}| < (7\times 10^{-13}/800~{\rm days})\times  (365/  0.39)
= 8\times 10^{-13} yr^{-1}
\end{equation}
%%%
%The corresponding rate of conversion for $\dot{M}/{M}$ is about several orders of magnitude smaller than the corresponding rate suggested in ref.\cite{mana}.
 
 We next proceed to compare these bound with the expectations of the
neutron$\to$ mixed star scenario.
%While this is not directly relevant to the bound, we also expect here
%that the energy to be emitted via neutrino pairs. The probability  
%$\sim G_F^2  K^4$  of emitting such a pair in $nn$ collisions with G-F and K
%the Fermi constant and the kinetic energy or temperature
%of neutrons or the neutrinos - is very small:  $\sim  10^{-34}$ for $K\sim KeV$ .
%%%%%%%%%%%%
%\color{red}{Why KeV?}

  The pulsars are observed while they are in the  electro-magnetically active ,``Beaming",  phase.
 It is precisely the fact that the   $\epsilon'\sim 10^{-11} -10^{-13}$ eV values
needed in order to achieve this while remaining consistent with all other
 bounds  on the $n\to n'$ mixing, is the main motivation for the work
of ref.\cite{mana}.

The authors of ref.\cite{mana}  start with an initial relatively broad pulsar
mass distribution with an average mass  of $1.6 M_{\odot}$. The final
mass distribution obtained is also broad and has an average of $1.25 M_{sun}$.
   However, at intermediate times when the heavier pulsars have
collapsed to black holes they obtain a much narrower distribution with
an average of  $1.35 M_{\odot}$, {\it if} the conversion of the lighter
pulsars is still ongoing.   Thus the best fit  in ~\cite{mana} is obtained if the
typical conversion times  are not just shorter than but{\it  comparable to} the pulsar lifetime in the range of $ 10^6-- 10^{10}$ years.   
%%%%%%%%%%%%%%%%%%%%%%%
%%%%%%%%%%%%%%%%%%%%%%%
%%%%%%%%%%%%%%%%%%%%%%%%%
  The corresponding mass loss rate in the Hulse-Taylor is
\begin{eqnarray}   \sim (0.3 \times 10 ^{-6} -  0.3\times  10^{-10}) M_{\odot}~ {yr}^{-1} \end{eqnarray}
 which exceeds the upper bound in Eq.8
by a factor of $ 10 - 10^{5}$.  The particular limit from the above two pulsars can be evaded, if the
neutron star to mixed star transition are shorter than their ages.  
%%%%%%%%%%%%%%%%
%%%%%%%%%%%%%%%%%%%%%
%%%%%%%%%%%%%%%%%%%%%%%%%%%
 
%%%%%%%
%%%%%%%%%

%\subsection{ Limits From Young Pulsars}

 This last possibility is however largely negated by observation of limits  from a particularly young star 
  PSR J1755-2550. This is  a young radio pulsar \cite{Ng et al.(2018)}. Discovered in 2015 and observed for 2.5 years, it is a member of a binary with orbital period of
 9.6963342(6) days. The spin down age is about $2.5\times 10^6$ years. The uncertainty in the measured orbital period over the 2.5 years time span of the observations implies a bound
 $|\frac{\dot P_b}{P_b}| \leq 2.1\times 10^{-12}$ per year. This is a young pulsar so that unlike PSR1913+16 one cannot argue that the conversion process was terminated long before the present epoch. This limit is 4 orders of magnitude smaller than the rate implied by the proposed conversion.
 
%\subsection{Limit from a Massive old neutron star}
 In general, even much less sensitive
 measurements  of period change of all
other pulsars in the approximately thirty binary systems known, would exclude them from being
candidates for the scenario of ref.\cite{mana}.
% where the best fit requires the
%pulsars to be still converting now.
 Clearly conversions of neutron stars to mixed stars can proceed at
rates much lower than those required to impact  the pulsar mass
distribution. This would still manifest in the orbital period change and
may allow us to exclude $\epsilon'$ values all the way down to
$10^{-15}e.V$ . Indeed in this case {\it all} pulsars would still be in the process of conversion to a
mixed star with increasing orbital period at the time of observation.

An important feature is that while ages, periods and period
stabilities greatly vary between the different pulsars, this is  NOT
the case for the roughly  constant length of the stellar conversion :
neutron star $\to$ mixed star, which  makes it increasingly difficult
for the conversion scenario to confront more and more binary pulsars.
 An example is the pulsar PSR J1614-2230 \cite{Demorest et al.(2010)}  which is a millisecond pulsar whose mass is $1.97 M_{\odot}$, in a binary system with a white dwarf companion of mass $0.5 M_{\odot}$. The pulse spin down age is 5.2 Gyr. Yet the pulsar has such a large mass. If the conversion did occur in   the past it must have started with an very large mass of $\sim 2.6 M_{\odot}$.  Most likely it would have collapsed to a black hole.
 
 Two other  pulsars which also allow us to set comparable constraints as above
 are: 
 
 (i) PSR J1141-6545~\cite{verbiest} which is a young pulsar with age 2 Myr
 with total mass $2.29 M_\odot$ ($M_{PSR}=1.27 M_\odot$ and $M_c=1.02M_\odot$. The residual rate of change of the binary orbital period (after taking care of the effect of ecceleration in the Galaxy as well as the kinematic effect and the expected gravitational radiation term)is $\dot{P_b}/P_b= - 7.8 \times 10^{-11} yr^{-1}$. 
It is a youmg pulsar so one cannot argue that the $nn'$ process has been already terminated. Moreover the sign is opposite from the predicted by the $nn'$ conversion.
 
 (ii) PSR J0437-4715 (age 1.6 Gyr) ~\cite{verbiest2}.    The residual rate of change of the binary orbital period (after taking care of the effect of ecceleration in the Galaxy as well as the kinematic effect) is  $|\dot{P_b}/P_b|= 8.2 \times 10^{-11}yr^{-1}$.  
 Then only caveat about (ii) is that it is a relatively old pulsar where $n-n'$ transition could have completed, although for the kind of limits on $\epsilon'$
 we derive from other pulsars listed, it would have taken longer than its age and we could still use this 
 to get a limit.

\subsection{Gravitational Waves Observations}
The recent observation (in gravitational waves and in much of the electromagnetic
spectrum) of a neutron star merger is most relevant to our discussion. In the
scenario of Ref.\cite{mana}, such mergers are likely to involve stars which are
already mixed. In this case the radii of the stars
should be considerably smaller than the $10 Km$ usually assumed. This then
causes the pattern of the emitted gravity waves to be different.
To appreciate the sensitivity to even moderate changes of radius - let
us consider the expected rate of GW emission . For approximately circular orbits, it is given
by:
\begin{equation}
\label{GW energy rate}
dW/dt =\frac{ 32 G_N}{5c^5} \mu^2\Omega_{orb} ^6 a^4 
\end{equation}
where  $\mu$ is the reduced mass: $\mu=(M_1M_2)/(M_1+M_2)$.
  Shrinking by the factor f the
radii of both stars  $R_1$ and $R_2$ will decrease the orbit size at the time of merging $a=R_1+R_2$
 by a same factor $f$. Using Kepler's  law $\Omega_{orb}  \sim  a^{-3/2}$, we find that
$dW/dt$ will increase by a factor of $f^{-5}$. There will also be reduction in $\mu$ which will tend to reduce
this increase. All these would lead 
to a considerable enhancement of the instantaneous gravity wave luminosity. For example for a reduction of radii $f\sim 0.7$, 
the enhancement will be about  four. %This will put stronger limits
%the $\epsilon'$. 
%To get the radius reduction factor $f$, we proceed as follows. Let us first adopt non-relativistic approximation and drop the Fermi energy contribution to the mass of the neutron star.
%The  mass of the pure neutron star then can be given as $M= m_b N -  \frac{3G_N N^2 m_b ^2}{ 5R}$ (where $m_b$ is the baryon mass and $N$ is the total baryon number). For the fully mixed star (with the same total baryon numbers) we have  $M_m= m_b N -  \frac{3G_N N^2 m_b ^2}{ 3R_m} $ ($M_m$ and $R_m$ refer to the mixed neutron star mass and radius). Given $M_m = 0.8M$ corresponding to the 20\% mass reduction in the scenario of ref.\cite{mana} and  noting that
%$3G_N N^2 m_b ^2/ 5R = 0.2 m_b N$, we get $f=R_m/R\simeq 0.55$ which gives $ f^{-5}\approx 19$,  Using this $f^{-5}$ and $\mu_{mixed}\simeq 0.8 M$ in Eq. 11, one gets a factor of $\sim 12$ increase in the instantaneous gravity wave luminosity. Even accounting for the Fermi energy, the factor will be larger than 10.
%While using Kepler's law and the above
%emission formula all the way down to the point of merging is a rough
%approximation, this still provides a sense of the drastic changes
%expected. 
In particular changes should occur in the predicted template that was fitted to the detailed  observed temporal  
GW signal and was consistent with the merging neutron stars being standard neutron stars.

The great advantage of this approach  is that unlike the one relying
on changing periodicities - the observation need  not be made while
the transition from the original neutron star to the mixed star is
ongoing. Indeed typically such mergers are expected to occur very late, during the cosmologically long period after
the individual pulsars have died.
{\it Furthermore the effect considered is maximal if the binary pulsar in question is old enough
so that the transition has terminated  ( or largely did so ) at the
time of observation and we have completely mixed stars with the
minimal radii possible.} Another constraint on the possible decrease of the neutron star radius has been obtained~\cite{new} from gravitational wave observations of the binary neutron star merger GW170817.   The authors obtained that the radius should be   be in the range $8.9\div 13.2 R_{\odot}$ with a mean value of $10.8 R_{\odot}$.

Thus this approach is somewhat complementary to the previous one and jointly
the two approaches tend to much more strongly restrict the scenario of
ref.\cite{mana}. Clearly one cannot, at the present  with such limited statistics
effectively use it to constrain the scenario of ref.\cite{mana} but this may
change in the future.

\section{Bounds on $\epsilon'$ from pulsar period increase measurements}
We next discuss  the bound on $\epsilon'$ which precision
measurements of pulsar periods can provide.
 We note that the above estimate (see Eq. (5)) in ref.\cite{mana} of, the rate of
$n\to{ n'}$ transitions in a neutron star described in~\cite{mana}
 [\ref{rate}] of $ 10^{-6} [\epsilon'/{10^{-11}eV}]^2~yr^{-1}$ may
be  somewhat optimistic.
This has no bearing on our main result- namely that the rate of
neutron star to mixed star transitions required for having an impact
on the pulsar mass distribution tend to conflict with upper bounds on
the  period increase of the orbital motion pulsars in the binary system.  It is
however relevant if we wish to use this approach to limit $\epsilon'$. In Appendix A,
we derive our estimate of $\Gamma_{nn'}$, which is somewhat lower:
\begin {eqnarray}
\Gamma_{n\to{n'}} =  0.6\times 10^{-7} [\epsilon'/{10^{-11} eV}]^2~{yr}^{-1}
\end{eqnarray}

%%%%%%%%%%%%%%%%%%%%%%%%
 % =   |1|$
%                    $|0|$
 %                                               $|\epsilon'|$
Using Eq. 8 and two estimates (optimistic and conservative), 
we find a bound on $\epsilon' \leq 10^{-15}$ eV or $10^{-13}$ eV respectively. We summarize our results in the following Table I using the estimate
 Eq. 12 (column 4) and that from Eq.6 (column (5))

\begin{table}[tbh]
\begin{center}
\begin{tabular}{|c||c||c||c||c|}\hline
Pulsar name & Age in yrs & Upper limit on & Upper limit on &Upper limit on  \\
& & $|\dot{M}/M|$ in ${yr}^{-1}$ & $\epsilon'$ in eV (using Eq. 12) &$\epsilon'$ in eV (using Eq.  6)\\\hline
PSR 1913+16 & $1.1\times 10^8$&$3\times 10^{-12} $& $7\times 10^{-14}$& $1.7\times 10^{-14}$\\\hline
PSR J1755+2550 & $2.1\times 10^6$ & $2.1\times 10^{-12}$ & $5.5\times 10^{-14}$ &$ 10^{-14}$\\\hline
PSRJ1952+2630 &$7.7\times 10^7$ & $8\times 10^{-13}$ & $3.5\times 10^{-14}$ & $\sim 10^{-14}$\\ \hline
PSR J1141-6545& $2\times 10^6$ & $3.4\times 10^{-11}$ &$2.2\times10^{-13}$ & $0.58\times 10^{-13}$\\\hline
PSR J0437-4715 & $1.6 \times  10^9$ & $4.1\times 10^{-11}$ & $3.2\times 10^{-13}$ & $1.2\times 10^{-13}$\\\hline\hline
\end{tabular}
\end{center}
\caption{Limits on $\epsilon'$ from several pulsar data using both the numbers of ref.Eq. 6 and our Eq. 12. We assume that 
the $n-n'$ conversion has not completed in these pulsars, in the spirit of ref.\cite{mana} where the best fit is obtained if the conversion is still ongoing. This is particularly unlikely for  PSR J1755+2550, whose age is only 2.1 Myr and for PSR J1141-6545 whose age is 2 Myr.}% and PSR J1141-6545 whose age is 1.4 Myr.}%if it did, that would require values of $\epsilon'$  considerably higher than what we have derived from period change constraints i.e. $\epsilon' \geq10^{-11}$ eV for PSR J1755+2550;  $\epsilon'\geq 2\times 10^{-12}$ eV and for PSR 1913+16 the lower limit is $\geq 0.6\times 10^{-12}$ eV. }%These limits are considerably higher than the limits we have derived from period change constraints.}
\end{table}

It is important to note that almost all discussions of $nn'$
oscillations and in particular the limits on $\epsilon' $ implied by
searches for such oscillations in neutron beams,  assume tiny
$\delta m'= m_n-m_{n'} = m-m'$ mass difference so that $ \delta m'
\leq {\epsilon'}$ and
$\delta m'\leq E_{magnetic} = \mu_{n}. B $(or $\mu_{n'}.B'$)  with $B
(B')$  the ordinary and mirror magnetic fields.
 If however the $\delta m'$ is (much) larger than both $\epsilon'$
and the magnetic energies, then $nn'$ oscillations would be suppressed
beyond detection. However the  $nn'$ transitions inside neutron stars
(which are suppressed by  much larger $\Delta E$ of order $20~ MeV$) are
insensitive to  $\delta m'$ so long as $\delta m'$ is smaller than
$\Delta E$ and the bounds  obtained in our analysis would extend to such
large $\delta m'$  values  %{ Need Figure of double log exclusion plain}

In passing we note that larger values of $\delta m'$  would be in
line with the comment made by two of us~\cite{MN}. We pointed out that the
mirror symmetry breaking at some high scale , which is required for
consistency with big bang nucleosynthesis %obtaining different  $\Omega(B')~ 5 \Omega(B)$ and different
with temperatures
$T'= (0.2-0.4)T$  tends to a "trickle down"  effect via loop diagrams,
barring some fine tuning, generates  $ \delta m'$ values much bigger
than what is usually assumed.

We further note that the proposal to understand the neutron decay anomaly by using neutron-mirror neutron oscillation requires $\epsilon'\simeq 10^{-10}$ eV~\cite{bere3} which is much larger than our upper bounds above. Our results would therefore rule out this explanation of neutron decay anomaly.

\section{ Some further comments and speculations}

It has been suggested that all ( or most) of
the ( super) heavy ,trans-lanthanide- elements are produced- by
ejecting neutron rich fragments into the host galaxy- in binary neutron star mergers. If this happens in 
completely mixed stars which contains equal amount of neutron and mirror neutrons, this would then imply similar abundance of ordinary and mirror
trans- Lantanide elements.
 More generally, the
microscopically exact mirror framework may also lead to proximity of
mirror and ordinary {\it atoms}. Some such proximity may be required if mirror matter is seeding the formation of galaxies.

%some
%$B'$ field is locally present and affects neutron to mirror neutron
%oscillations in neutron beam experimental set-ups.

This brings up the more general question which we will briefly touch
upon, of how much mirror matter is expected in the galaxy, the solar
system and in earth ?
The different $\Omega'$ and $T'$ in the mirror and ordinary sector
causes a much more abundant mirror helium He' production at
nucleosynthesis in the mirror sector~\cite{bere}. This in turn leads to a  very
different pattern of star formation, supernovae and element abundance
which generally is expected to be shifted to heavier elements in the
mirror sector.
 At least $80$ percent of the mirror baryonic mass should be in the
form of massive stars in elliptical galaxies as otherwise the large
$He'-He'$ scattering cross-sections -identical to those of $ He-He$ of $\approx 
10^{-16} cm^2$ -would dramatically violate the upper bounds suggested
by the bullet cluster.
 Still since the mirror matter is dissipative  we expect that it  will tend to co-cluster with ordinary matter due to their mutual gravity. Thus, it
  can be searched for
 in earth, on the lunar surface, and in meteorites.
 Conversely aggregates of mirror matter such as micro-haloes and
mirror matter stars should include some small amount of ordinary
matter - which in turn may dramatically change their properties in an
observable manner.%{We thank Jim Buckley for pointing this out and for useful discussions.}.

\section{Summary and conclusion} In conclusion, in this paper, we have noted that astrophysical
data pertinent to precision pulsar timing and binary neutron star merger can be used to restrict the scenario of ref.\cite{mana} where $n-n'$ mixing causes transition from a pure neutron star to mixed neutron stars. In turn, this allows us to restrict particle physics
parameters such as the mixing between neutron and mirror neutron ($n\to n'$) which  possible in an almost exact mirror symmetric dark sector. Our conclusion is based on the 
 recently proposed scenario
~\cite{mana} of neutron star to mixed neutron star transition which can arise due to $n\to n'$ transition. We find that the key parameter responsible for $n\to n'$ transition
$\epsilon'$ is restricted to be below $10^{-13}$ eV to $10^{-14}$ eV by current pulsar timing data. While the constraints from the binary pulsars of  age $\sim 100$ Myr could be evaded by assuming that the transition has already been  completed,  it is much more difficult to do so for the two pulsars with age of
$\sim 2$ Myr. Also the case of the old massive ( 2 solar masses) pulsar casts doubt on the scenario  of mass reduction,
The observations of the gravitational waves from the the
binary neutron star merger GW170817 constrains a reduction of the neutron stars radii which is implied by the $nn'$ process. Additionally, the constraint from the merger of 2 neutron stars detected by LIGO and VIRGO as well as by gamma ray observation is not sensitive to the time scale of the transition as these stars are likely old ones.  

An interesting contrast between the current  pulsar timing limits and the limits that can be obtained from laboratory searches is that our limit is valid for $n-n'$ mass differences  $\delta m'$ as large as 20 MeV whereas the latter limits are valid only for much smaller splittings such as those caused by the local magnetic field difference  between the visible and the mirror world.

%We also suggest how these limits could be improved by gravitational wave observations from binary neutron star mergers.

 %%%%%%%%%%%%%%%
\section{Acknowledgement} We thank Mark Alford, Jim Buckley and Richard Mushotsky for extensive discussions and one of us (S. N.) would like to thank  Anca Turneau and David Milstead for the invitation to the 2018 ESS neutron meeting. The work of R. N. M. is supported by the US National Science Foundation under Grant No. PHY1620074. The work of I. G. was supported  by the Afeka Research Fund. 

%\section*{Appendix A}
\appendix

\section{}
Two factors enter the above estimate :$ \Gamma( n\rightarrow{n'}) = 
\Gamma(nn)P_{nn'}$:
the probability of having an $n'$ at the time of $nn$ collision $P_{nn'}$, and the
rate of such collisions $\Gamma_{nn}$.  This reflects a simple,
physically motivated picture~\cite{cowsik} in which
one assumes that:
a)  The coherent build up of  $|n'>$ in the initial purely $|n>$ state
of the two component system proceeds unimpeded by nuclear interactions
during the time of flight between two consecutive collisions .
b) The coherent build-up stops upon collision and the $n'$ part is
released as out-going mirror neutron particles.

%However both factors seem to have been over-estimated in \cite{mana}.

 We note that the high $nn$
collision rate in Eq. \ref{value} implies a very short flight time
\begin{eqnarray}
    t_{nn} = 1/\Gamma_{nn} \approx {( 3-4) \times10^{-25} sec}.    \label{flighttime}
\end{eqnarray}
separating consecutive collisions.

A key point is that during flight times $t_{nn}$ shorter than
$\Delta E ^{-1}$, the admixture of  the
mirror neutron $|n'>$  does $\it{not}$ build up to it's asymptotic
value of $\epsilon'/ {\Delta E}$  which was implicitly used in reference \cite{mana} to
estimate $P_{nn'}=[\epsilon'/ {\Delta(E)}]^2 $.  

The free evolution of the
initial pure neutron state during the short time of flight starting with
$\psi(0)=|n>=\left(\begin{array}{c}1\\0\end{array}\right)$   yields $\psi(t)>=e^{( iHt)}|\psi(0)> \approx (1+
iHt)|\psi(0)>\approx\left(\begin{array}{c}1\\\epsilon' t\end{array}\right)$ %where $H=\Delta E\sigma_z+\epsion'\sigma_x$ is the Hamiltonian in

where $H = \Delta(E) \sigma_z + \epsilon'\sigma_x$  is the
Hamiltonian in the two dimensional $|n>,|n'>$ Hilbert space.
 This
then yields a probability of generating a mirror neutron $n'$ between
two consecutive collisions  of 
$P_{nn'} =[\epsilon'. t_{nn}]^2)$.
%  which
%is smaller than
% $P_{old}  =[ \epsilon'/(E(F)-E'(F)]^2$ used above~\cite{mana} by $ t_{nn}
%E(F)]^2 \approx {10^{-3}}$  for $E(F)= 40 MeV$.

Substituting the new $P_{nn'}$ and $\Gamma_{nn} = t_{nn}^{-1} $ in $
\Gamma_{n\to(n')}= \Gamma_{nn}.P $ we find an alternative
expression ( appropriate for short $t_{nn}$ )
\begin{eqnarray}
    \Gamma_{n\to n'} = t_{nn}{\epsilon'}^2       \label{new}
    \end{eqnarray}
 Using $ t_{nn}\approx {10^{-23}} ~sec$ - the time required to travel the]
$O(Fermi)$ distance between neighboring neutrons at $1/3$ of  the
speed of light  then leads to a new estimate of the  conversion rate:
\begin {eqnarray}
\Gamma_{n\to{n'}} =  0.6\times 10^{-7} [\epsilon'/{10^{-11} eV}]^2~{yr}^{-1}
\end{eqnarray}
Which is about 20 times smaller than the estimate of ref.\cite{mana} . 
Clearly this
is a very rough estimate. In particular having the nuclear medium
manifest just as a series of frequent collisions is very  crude. The
conversion is ongoing all the time and the proper treatment should use
the Schrodinger equation in the medium as in the careful
treatment of  $n\rightarrow \bar{n}$ nuclear transitions by ~\cite{Richard}.
 The latter works
yield transition rates somewhat faster than those in \cite{cowsik}.
 In mapping out the upper limits on $\epsilon'$ in the Table I, we use
both this estimate and the
original estimate of ref.\cite{mana}) in Eq.6.


\begin{thebibliography}{99}

 \bibitem{mirror1} T. D. Lee and C. N. Yang, Phys. Rev. 104, 254 (1956); K. Nishijima, private communication; Y. Kobzarev, L. Okun and I. Ya Pomeranchuk, Yad. Fiz. 3, 1154 (1966); M. Pavsic, Int. J. T. P. 9, 229 (1974); S. I. Blinnikov and M. Y. Khlopov, Astro. Zh. 60, 632 (1983); E. W. Kolb, D. Seckel and M. Turner, Nature, 514, 415 (1985); R. Foot, H. Lew and R. Volkas, Phys. Lett. B 272, 67 (1991);  
 
  \bibitem{mirror3} R. Foot and R. Volkas, Phys. Rev. D 52, 6595 (1995);  
 
 \bibitem{mirror2} Z. Berezhiani and R. N. Mohapatra, Phys. Rev. D 52, 6607 (1995).
 
 
  \bibitem{asym} Z. Berezhiani, A. Dolgov and R. N. Mohapatra, Phys. Lett. B 375, 26 (1996).
  
  \bibitem{MN} R.~N.~Mohapatra and S.~Nussinov,
  %``Constraints on Mirror Models of Dark Matter from Observable Neutron-Mirror Neutron Oscillation,''
  Phys.\ Lett.\ B {\bf 776}, 22 (2018)
 % doi:10.1016/j.physletb.2017.11.022
  [arXiv:1709.01637 [hep-ph]].
  
  
  
  \bibitem{bere} Z.~Berezhiani, P.~Ciarcelluti, D.~Comelli and F.~L.~Villante,
  %``Structure formation with mirror dark matter: CMB and LSS,''
  Int.\ J.\ Mod.\ Phys.\ D {\bf 14}, 107 (2005); Z.~Berezhiani, D.~Comelli and F.~L.~Villante,
  %``The Early mirror universe: Inflation, baryogenesis, nucleosynthesis and dark matter,''
  Phys.\ Lett.\ B {\bf 503}, 362 (2001); A.~Y.~Ignatiev and R.~R.~Volkas,
  %``Mirror dark matter and large scale structure,''
  Phys.\ Rev.\ D {\bf 68}, 023518 (2003).
  
    \bibitem{foot} For a review and references, see R.~Foot,
  %``Mirror dark matter: Cosmology, galaxy structure and direct detection,''
  Int.\ J.\ Mod.\ Phys.\ A {\bf 29}, 1430013 (2014); Z.~Berezhiani,
  %``Through the looking-glass: Alice's adventures in mirror world,''
  In *Shifman, M. (ed.) et al.: From fields to strings, vol. 3* 2147-2195;  [hep-ph/0508233].
 

    
      
    \bibitem{BB} Z.~Berezhiani and L.~Bento,
  %``Neutron - mirror neutron oscillations: How fast might they be?,''
  Phys.\ Rev.\ Lett.\  {\bf 96}, 081801 (2006); Z.~Berezhiani,
  %``More about neutron - mirror neutron oscillation,''
  Eur.\ Phys.\ J.\ C {\bf 64}, 421 (2009).
  
   \bibitem{expt} A.P. Serebrov et al., Phys. Lett. B 663, 181 (2008); I. Altarev et al., Phys. Rev. D 80, 032003 (2009); V.~Bondar [nEDM at PSI Collaboration],
  %``Searches for exotic interactions with the nEDM experiment,''
    arXiv:1607.07293 [hep-ph].
    
    \bibitem{yuri} L. Broussard and Y. Kamyshkov, Talk at BLV 2017; Z.~Berezhiani, M.~Frost, Y.~Kamyshkov, B.~Rybolt and L.~Varriano,
  %``Neutron Disappearance and Regeneration from Mirror State,''
  Phys.\ Rev.\ D {\bf 96}, 035039 (2017).
  
  \bibitem{nnbar} V.~A.~Kuzmin,
  %``Cp violation and baryon asymmetry of the universe,''
  Pisma Zh.\ Eksp.\ Teor.\ Fiz.\  {\bf 12}, 335 (1970);
  S.~L.~Glashow,
  %``The Future of Elementary Particle Physics,''
  NATO Sci.\ Ser.\ B {\bf 61}, 687 (1980); R.~N.~Mohapatra and R.~E.~Marshak,
  %``Local B-L Symmetry of Electroweak Interactions, Majorana Neutrinos and Neutron Oscillations,''
  Phys.\ Rev.\ Lett.\  {\bf 44}, 1316 (1980)
  Erratum: [Phys.\ Rev.\ Lett.\  {\bf 44}, 1643 (1980)].

\bibitem{review} For recent reviews, see R.~N.~Mohapatra,
  %``Neutron-Anti-Neutron Oscillation: Theory and Phenomenology,''
  J.\ Phys.\ G {\bf 36}, 104006 (2009)
  %doi:10.1088/0954-3899/36/10/104006
  [arXiv:0902.0834 [hep-ph]]; D.~G.~Phillips, II {\it et al.},
  %``Neutron-Antineutron Oscillations: Theoretical Status and Experimental Prospects,''
  Phys.\ Rept.\  {\bf 612}, 1 (2016)
  %doi:10.1016/j.physrep.2015.11.001
  [arXiv:1410.1100 [hep-ex]].
  
  \bibitem{baldo}   M. Baldo-Ceolin et al, Z. Phys. C63, 409 (1994)
 
  
  \bibitem{superk} J.~Gustafson {\it et al.} [Super-Kamiokande Collaboration],
  %``Search for dinucleon decay into pions at Super-Kamiokande,''
  Phys.\ Rev.\ D {\bf 91}, no. 7, 072009 (2015)
 % doi:10.1103/PhysRevD.91.072009
  [arXiv:1504.01041 [hep-ex]]; B.~Aharmim {\it et al.} [SNO Collaboration],
  %``Search for neutron-antineutron oscillations at the Sudbury Neutrino Observatory,''
  Phys.\ Rev.\ D {\bf 96}, no. 9, 092005 (2017)
 % doi:10.1103/PhysRevD.96.092005
  [arXiv:1705.00696 [hep-ex]].
  
 
  
  
 
\bibitem{mana} M. Mannarelli, Z. berezhiani, R. Biondi and F. Tonnelli, Talk at the Nordita ESS workshop, December (2018).

\bibitem{n*} J. Lattimer, Ann. Rev. Nucl. Part. Sci. {\bf 62}, 485 (2012).

\bibitem{GN} I.~Goldman and S.~Nussinov,
  %``Limit on Continuous Neutrino Emission from Neutron Stars,''
  JHEP {\bf 1008}, 091 (2010)
 % doi:10.1007/JHEP08(2010)091
  [arXiv:0907.1555 [astro-ph.SR]].
  

  
  \bibitem{GMNRT} I.~Goldman, R.~N.~Mohapatra, S.~Nussinov, D.~Rosenbaum and V.~Teplitz,
  %``Possible Implications of Asymmetric Fermionic Dark Matter for Neutron Stars,''
  Phys.\ Lett.\ B {\bf 725}, 200 (2013)
 % doi:10.1016/j.physletb.2013.07.017
  [arXiv:1305.6908 [astro-ph.CO]].
 
 \bibitem{OTHER} See for example, C.~Kouvaris and N.~G.~Nielsen,
  %``Asymmetric Dark Matter Stars,''
  Phys.\ Rev.\ D {\bf 92}, no. 6, 063526 (2015)
  %doi:10.1103/PhysRevD.92.063526
  [arXiv:1507.00959 [hep-ph]]; P.~Ciarcelluti and F.~Sandin,
  %``Have neutron stars a dark matter core?,''
  Phys.\ Lett.\ B {\bf 695}, 19 (2011)
 % doi:10.1016/j.physletb.2010.11.021
  [arXiv:1005.0857 [astro-ph.HE]]; J.~Ellis, G.~Hütsi, K.~Kannike, L.~Marzola, M.~Raidal and V.~Vaskonen,
  %``Dark Matter Effects On Neutron Star Properties,''
  Phys.\ Rev.\ D {\bf 97}, no. 12, 123007 (2018)
  %doi:10.1103/PhysRevD.97.123007
  [arXiv:1804.01418 [astro-ph.CO]].
  \bibitem{Alford} M. Alford, S. Han and M. Prakash,
 JPS .conf. proc   013041(2014); 
  \bibitem{Bedaque and Steiner} P. Bedaque and A.W. Steiner, Phys. Rev. Lett 114   No 3 . 031103 ( 2015)
    \bibitem{jeans} J. H. Jeans, MNRAS {\bf 85}, 2 (1924).
 


  \bibitem{Demorest et al.(2010)}  Demorest, P.~B., Pennucci, T., Ransom, S.~M., Roberts, M.~S.~E., \& Hessels, J.~W.~T. \nat, {\bf 467}, 1081 (2010)

  \bibitem{Ng et al.(2018)}  Ng, C., Kruckow, M.~U., Tauris, T.~M., et al.   MNRAS, {\bf 476}, 4315  (2018)
  
  \bibitem{Lazarus et al. 2014}  Lazarus P., et al.,  MNRAS, {\bf 437}, 1485 (2014).
  
  \bibitem{verbiest} N.~D.~R.~Bhat, M.~Bailes and J.~P.~W.~Verbiest,
  %``Gravitational-radiation losses from the pulsar-white-dwarf binary PSR J1141-6545,''
  Phys.\ Rev.\ D {\bf 77}, 124017 (2008).

  
  \bibitem{verbiest2} J.~P.~W.~Verbiest {\it et al.},
  %``Precision timing of PSR J0437-4715: an accurate pulsar distance, a high pulsar mass and a limit on the variation of Newton's gravitational constant,''
  Astrophys.\ J.\  {\bf 679}, 675 (2008).
 
  
  \bibitem{nano} M. T. Lam et al. arXiv: 1809.03058 (Astro-ph);  M. A. McLaughlin, {\rm Classical and Quantum Gravity} {\bf 30}, issue 22, id. 224008 (2013).
  
  \bibitem{new} S.~De, D.~Finstad, J.~M.~Lattimer, D.~A.~Brown, E.~Berger and C.~M.~Biwer,
  %``Tidal Deformabilities and Radii of Neutron Stars from the Observation of GW170817,''
  Phys.\ Rev.\ Lett.\  {\bf 121}, no. 9, 091102 (2018)
  Erratum: [Phys.\ Rev.\ Lett.\  {\bf 121}, no. 25, 259902 (2018)]

  
   \bibitem{bere3} Z. Berezhiani, arXiv:1807.07906.
   
    \bibitem{cowsik} R.~Cowsik and S.~Nussinov,
  %``SOME CONSTRAINTS ON DELTA B = 2 n anti-n OSCILLATIONS,''
  Phys.\ Lett.\  {\bf 101B}, 237 (1981).
 
 \bibitem{Richard} C.~B.~Dover, A.~Gal and J.~M.~Richard,
  %``Neutron Anti-neutron Oscillations In Nuclei,''
  Phys.\ Rev.\ D {\bf 27}, 1090 (1983).
  

  
  \end{thebibliography}
\end{document}